\documentclass[twocolumn,prb,showpacs,10pt]{revtex4}
\usepackage{psfrag}
\usepackage{graphicx}
\usepackage{amsmath}
\usepackage{amssymb}
\usepackage{subfig}

\usepackage{color}
\usepackage{multirow}
\begin{document}
\title{A density functional theory based direct comparison of coherent tunnelling and electron hopping in
redox-active single molecule junctions}

\author{Georg Kastlunger and Robert Stadler}
\affiliation{Department of Physical Chemistry, University of Vienna, Sensengasse 8/7, A-1090 Vienna, Austria} 
\affiliation{\mbox{Institute for Theoretical Physics, Vienna University of Technology,}\\ Wiedner Hauptstrasse 8-10, A-1040 Vienna, Austria\\ 
Email: robert.stadler@tuwien.ac.at}

\date{\today}

\begin{abstract}
For defining the conductance of single molecule junctions with a redox functionality in an electrochemical cell, two conceptually different electron transport mechanisms, namely coherent tunnelling and vibrationally induced hopping compete with each other, where implicit parameters of the setup such as the length of the molecule and the applied gate voltage decide which mechanism is the dominant one. Although coherent tunnelling is most efficiently described within Landauer theory, while the common theoretical treatment of electron hopping is based on Marcus theory, both theories are adequate for the processes they describe without introducing accuracy limiting approximations. For a direct comparison, however, it has to be ensured that the crucial quantities obtained from electronic structure calculations, i.e. the transmission function T(E) in Landauer theory, and the transfer integral \textit{V}, the reorganisation energy $\lambda$ and the driving force $\Delta G^0$ in Marcus theory, are derived from similar grounds as pointed out by Nitzan and co-workers in a series of publications. In this article our framework is a single particle picture, where we perform density functional theory calculations for the conductance corresponding to both transport mechanisms for junctions with the central molecule containing one, two or three Ruthenium centers, respectively, where we extrapolate our results in order to define the critical length for the transition point of the two regimes which we identify at 5.76 nm for this type of molecular wire. We also discuss trends in dependence on an electrochemically induced gate potential.
\end{abstract}
\pacs{73.63.Rt, 73.20.Hb, 73.40.Gk}
\maketitle

\begin{section}{Introduction}\label{sec:intro}

\begin{figure*}
\includegraphics[width=\textwidth,angle=0]{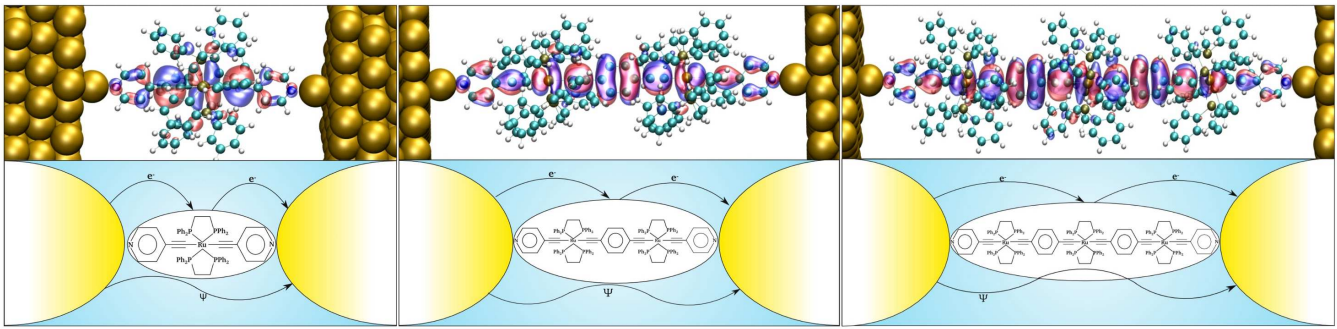}
\caption[cap.Pt6]{\label{fig.structures}Atomic structures (upper panels) and chemical formulae (lower panels) of Ruthenium bis(pyridylacetylide) complexes with one, two, and three Ru(PPh$_{2}$)$_{4}$(C$_{2}$H$_{4}$)$^{2}$-centers, respectively, which are coupled to gold leads via pyridil-anchor groups. For all junctions in this figure NEGF-DFT calculations for coherent tunneling (as indicated by the wiggled line below the formulae) as well as DFT calculations of the 2-step electron hopping process following the recipe given in the main text (the two arrows above the formulae) have been carried out. In the upper panels, the respective HOMOs for all structures, which are crucial in both transport regimes, are also shown on top of the molecular structures.}
\end{figure*}

Electron transport through single molecule junctions in ultra high vacuum (UHV) and temperatures is commonly described with a nonequilibrium Green's function (NEGF) approach~\cite{keldysh} in combination with a density functional theory (DFT) based description of the electronic structure of the leads and the scattering region of the junction~\cite{atk}$^{-}$\cite{kristian}. The modelling of the conductance and current/voltage (I/V) characteristics of single molecules at ambient conditions, at which an electrochemical scanning tunneling microscope (STM)~\cite{Tim1,Tim2,Nichols1,Nichols2,Evers} is operating and which are necessary for the practicability of devices, is more challenging, because here two competing electron transport mechanisms have to be considered, namely electron hopping and coherent tunneling where it depends on the adjustment of an electrochemical gate voltage as well as on structural details of the system which mechanism dominates the accumulated transfer rate of electrons. The distinction between the two is important for the design of molecular wires, where coherent tunneling prevails at short length regardless of the chemical structure but decays exponentially and then at some structure dependent crossover point in molecular length gives way to the Ohmic behavior of electron hopping, which is crucial for making a wire of any use in real life~\cite{ratner3}. The application of a gate potential for e.g. implementing transistor properties~\cite{Tim3} or for optimizing the conductance properties of a wire is also easier to achieve in an electrochemical setup where no third electrode has to be placed close to the leads for source and drain and no strong local electric fields are required.~\cite{Tim4}

There have been a variety of experiments aiming at a direct detection of the crossover length in molecular wires, where coherent tunneling is replaced by hopping. In a pioneering series of papers Ratner, Wasielewski and co-workers~\cite{ratner1}-~\cite{ratner2} investigated the intramolecular electron transfer from a donor to an acceptor moiety via a bridge which consisted of oligo phenylene-vinylene (OPV) molecules of increasing length, where optical absorption spectra allowed the derivation of charge separation and recombination rates and a switch in the transfer mechanism was detected when the bridge consisted of more than two monomers. More recently, Choi et al. measured the conductance of monolayers of oligophenyleneimine (OPI) thiolates adsorbed on a gold substrate with their length varying between 1.5 and 7 nm by using a STM and found the transition point at $\sim$ 4nm~\cite{frisbie1}, while single molecule measurements on the corresponding dithiolates in a break junction setup placed this crossover point in the range of 5.2-7.3 nm~\cite{hines}. Lu et al. studied monolayers of oligo phenylene-ethynylene (OPE) wires, where a transition from tunneling to hopping was observed at a molecular length of $\sim$ 2.75 nm~\cite{wang1}, while a transition length of about 3 nm was found for a series of oligo arylene ethynylene (OAE) derivatives in single molecule measurements by Wandlowski and co-workers~\cite{wandlowski1}.

All theoretical attempts to identify and characterize the length for the transition between tunneling and hopping so far~\cite{ratner5}-~\cite{nitzan3} to our best knowledge suffer from two severe limitations which are also related to each other: i) Hopping is characterized by model or tight binding Hamiltonians where crucial parameters such as onsite energies of monomer sites or the coupling between them are just set to some reasonable values and not derived from ab initio or any other type of electronic structure calculations, which would reflect their dependence on any details of the molecular structure under investigation. ii) As a consequence only N-step hopping could be investigated where it is assumed that the electron hops from one monomer to the other until it reaches the final Nth one. This is a reasonable assumption for simulations on DNA wires, on which indeed the focus of the theoretical articles quoted above was, but the assumption is not justified at all for the highly conjugated OPV, OPI, OPE and OAE wires which have been in the recent experimental spotlight due to their higher conductance and neither for the wires with redox active centers which we describe further below. For conjugated systems in general one expects 2-step hopping~\cite{ratner2} where an electron hops from the donor (or left lead) to the bridge (or central molecule) in a first step and then on to the acceptor (or right lead) in a second one.

While coherent tunneling is nowadays routinely described within the single-particle framework of NEGF-DFT~\cite{atk}$^{-}$~\cite{kristian}, where we have recently shown that the oxidation state of the redox active center in the scattering region can also be adjusted in two different ways~\cite{first}, no ab initio procedure for the description of electron hopping in single molecule junctions has so far been proposed. Our understanding of the hopping process in general is relying on Marcus' theory of electron transfer reactions~\cite{marcus1}-~\cite{marcus3}, which are the rate limiting step in many redox reactions and where the activation can be achieved thermally, photo-chemically or by applying an external potential. The key parameters in this theory, namely, the transfer integral \textit{V}, the reorganization energy $\lambda$ and the driving force $\Delta G^{0}$ are usually derived from quantum chemical techniques for intra-~\cite{jerome1} and inter-molecular electron transfer~\cite{jerome2} but for their definition and derivation in the context of a single molecule junction, a variety of technical as well as more fundamental issues arise, which to address is one of the main achievements of the work we present in this article.

We carried out DFT calculations for both electron transport regimes, namely coherent tunneling and hopping, for the three single molecule junctions depicted in Figure\ref{fig.structures}. We chose these particular type of molecules due to their high all-through conjugation and therefore also high conductivity as well as the presence of redox-active Ru centers which allow for efficient electro-chemical gating. They also offer the possibility to alter their local redox state independently, which would require some structural modifications but this is not the topic of our current article. Because of these benign properties of the displayed Ru complexes, conductance measurements and optical absorption experiments have been carried out on them where thiol~\cite{wang2} and cyano anchors~\cite{frisbie2} have been used instead of the pyridil linkers in our article which we prefer due to their stability under ambient conditions, high junction formation probability and because they do not require protecting groups during the adsorption process on the electrodes~\cite{wandlowski1}. 

The paper is organized as follows: In the next section we give a detailed account of our theoretical framework for the two different electron transport regimes with an emphasis on how to obtain crucial parameters and quantities from DFT. In the following sections we use these methods for deriving the crossover point in the molecular length dependence of the conductance for the systems in Figure\ref{fig.structures} from first principles and also discuss the effect of an electrochemical gate potential. In the last section we provide a summary of our results.

\end{section}

\begin{section}{Theoretical Framework and Computational Details}

Up to now the theoretical understanding of electron hopping in single molecule junctions has been driven by the phenomenological models of Kuznetsov and Ulstrup~\cite{ulstrup1}-~\cite{ulstrup4} and some earlier work by Schmickler~\cite{schmickler}, while more recently Nitzan addressed the relation between the conductance as the quantity calculated by the NEGF-DFT formalism for coherent tunneling and the transfer rate in Marcus' theory of electron hopping from a formal perspective~\cite{nitzan4}-~\cite{nitzan5}. Nitzan and Migliore also developed a single particle framework for 2-step hopping~\cite{nitzan6}, which is distinct from the usual picture based on enthalpies and total energies and therefore allows for an orbital interpretation. We make use of this framework heavily in our work as is discussed further below but while in Ref.~\cite{nitzan6} general formal relationships are established and some typical values of \textit{V}, $\lambda$ and $\Delta G^{0}$ are used as a means of illustration without referring to a particular molecular system, the aim in our work is to derive these quantities for the junctions in Figure~\ref{fig.structures} from first principles. In Marcus theory the electron transfer process is described as a chemical reaction, where in our specific case it is an oxidation of the ground state of the molecules which in a quantum chemical picture corresponds to the removal of an electron from the HOMO.

In our modelling of electron hopping we assumed a 2-step process, where a positive charge jumps from the left lead to the highest occupied molecular orbital (HOMO) of the molecule, which we know to be conjugated throughout the whole molecule from our previous work~\cite{first}-~\cite{second}, and then from there on to the right lead in a second step. Ratner and co-workers have pointed out that the situation might get more complicated even in conjugated wires due to torsional degrees of freedom, and electron transfer rates might be affected by their thermal activation~\cite{ratner2,ratner4}. While the phenyl groups separating two adjacent Ru centers in the structures shown in Figure\ref{fig.structures} are quite efficiently trapped by steric repulsion to the bulky substituents on the Ruthenium, the pyridil anchors have indeed some flexibility which could break the conjugation ranging otherwise from the left lead to the right one. We performed DFT based total energy calculations, which showed that the energy barriers are 274 and 90 meV for rotating the phenyl ring in the middle and the pyridil groups at the ends of the molecule, respectively. Since room temperature corresponds to only 25 meV, we can still assume that the conjugation is more or less undisturbed at ambient conditions if present from the beginning, i.e.  if conformeres have been successfully separated after chemical synthesis.

We calculated G$_{hop}$ for room temperature and G$_{coh}$ for 0 Kelvin. This might seem counterintuitive since G$_{hop}$ depends considerably on the temperature because this type of electron transport is thermally activated or in other words at 0 K there is no conductance due to hopping. By contrast, G$_{coh}$ does not depend on the temperature in a first order approximation, but is only affected by electron-phonon coupling as a second order effect. So our assumption was that we model room temperature behaviour, which we treat explicitly for the hopping but where we neglect thermal effects for coherent tunneling.

\begin{subsection}{Electronic Structure Calculations}

All electronic structure calculations in this article were performed with the GPAW code~\cite{GPAW1,GPAW2}, where the core electrons are described with the projector augmented wave (PAW) method and the basis set for the Kohn Sham wavefunctions has been chosen to be a linear combination of atomic orbitals (LCAO) on a double zeta level with polarisation functions (DZP) for all electronic structure calculations. The sampling of the potential energy term in the Hamiltonian is done on a real space grid when using GPAW, where we chose 0.18 \AA{} for its spacing and a Perdew-Burke-Ernzerhof (PBE)~\cite{PBE} parametrisation for the exchange-correlation (XC) functional throughout this article. 

\end{subsection}

\begin{subsection}{Coherent Tunneling}

Within NEGF-DFT~\cite{atk}$^{-}$~\cite{kristian} the transmission function T(E) for coherent tunneling is defined by $T(E)=Tr(G_{d}\Gamma_{L}G_{d}^{\dagger}\Gamma_{R})$ where $G_{d}=(E-H_{d}-\Sigma_{L}-\Sigma_{R})^{-1}$ represents the Greens function of the device containing the self energy matrices $\Sigma_{L/R}$ due to the left/right lead, $\Gamma_{L/R}=i(\Sigma_{L/R}-\Sigma_{L/R}^{\dagger})$ and H$_{d}$ the Hamiltonian matrix for the device region, which contains not only the Ru-complex but also 3-4 layers of the aligned Au surface on each side. Due to the rather large size of the central molecules, we had to use gold slabs with a 6x6 unit cell in the surface plane in order to ensure that neighbouring molecules do not interact. For the same reason we used a 2x2x1 {\bf k} point grid corresponding to only 2 k points in the irreducible Brillouin zone for all transmission functions discussed in this article. 

\end{subsection}

\begin{subsection}{Electron Hopping}

In order to describe electron hopping in single molecule junctions, the famous Marcus Hush formula~\cite{marcus2,hush,jerome3} for the transfer rate in intra-or inter-molecular electron transfer reactions needs to be modified because the initial and final states have to be replaced by the manifold of all occupied and unoccupied surface states in the lead with the right symmetry. This was first recognized by Chidsey~\cite{chidsey}, who modified the Marcus Hush formula by introducing an integral over all metal states. In the present case we are dealing with a 2-step reaction, where in the first step we oxidize the molecule by charging it with a hole which is supposed to come from the left lead, and in the second step decharge the molecule again, i.e. reduce it, where the corresponding hole moves on to the right electrode. The overall conduction process is then described by the transfer rates of the corresponding two separate electron transfer reactions: 

\begin{equation}
k_{ox}=\frac{2\pi}{\hbar}V^{2}\frac{1}{\sqrt{4\pi\lambda k_{b}T}}\int e^{-\frac{(\lambda+\Delta G^{0}+\epsilon)^{2}}{4\lambda k_{b}T}}(1-f(\epsilon))d\epsilon\label{eqn.kox}
\end{equation}
\begin{equation}
k_{red}=\frac{2\pi}{\hbar}V^{2}\frac{1}{\sqrt{4\pi\lambda k_{b}T}}\int e^{-\frac{(\lambda-\Delta G^{0}+\epsilon)^{2}}{4\lambda k_{b}T}}f(\epsilon)d\epsilon\label{eqn.kred}
\end{equation}

where $f(\epsilon)=1/(e^{\epsilon/(k_{b}T)}+1)$ is the Fermi function, which is part of both equations because in the oxidation reaction only the unoccupied states of the left lead can provide a positive charge (hole) and only the occupied states of the right lead can absorb it in the reduction of the molecule, where the thermal broadening of the Fermi levels of the leads at finite temperatures is also built into $f(\epsilon)$.

For calculating the overall conductance of the junction in the hopping regime, Migliore and Nitzan derived an expression containing both $k_{ox}$ and $k_{red}$ for including the effect of both steps of the process~\cite{nitzan6}, which in our symmetric case with the same electrode material and surface orientation as well as the same anchor group on both sides of the junction becomes: 

\begin{equation}
G_{hop}=\frac{e^{2}}{2k_{b}T}\frac{k_{ox}k_{red}}{k_{ox}+k_{red}}\label{eqn.g_hop}
\end{equation}

where the quantities $V$, $\lambda$ and $\Delta G^{0}$ need to be defined on a single particle level in order to be able to make a direct comparison with the conductance for coherent tunneling G$_{coh}$ which can be simply obtained by taking the value of the transmission function $T(E)$ as computed with NEGF-DFT at the Fermi energy. We note that Equation~\ref{eqn.g_hop} applies only in the limit of zero bias, i.e. for an infinitesimally low potential difference between the source and drain electrodes. It is, however, fully applicable for finite electrochemical gate voltages, where the respective potential is just added to the value of $\Delta G^{0}$ in equations~\ref{eqn.kox} and~\ref{eqn.kred}. 

\begin{subsubsection}{Transfer Integral}

In a recent article we have shown that the transfer integral \textit{V$_{Au-Au}$} between the metal electrodes bridged by the molecule can be also used to determine the conductance for coherent tunneling~\cite{second}. In the context of electron hopping, however, the conductance is defined by two consecutive reactions, where for both another transfer integral \textit{V$_{Au-Mol}$} is needed which describes the electronic coupling between the molecule and one of the leads. In contrast to \textit{V$_{Au-Au}$} which we calculated only at the Fermi Energy E$_{F}$ in order to define G$_{coh}$, for electron hopping we need \textit{V$_{Au-Mol}$} to be integrated over all energies. This information can be neatly retrieved from the peak in $T(E)$ corresponding to the HOMO as calculated with NEGF-DFT for coherent tunneling, because the width of this peak and \textit{V$_{Au-Mol}$} are directly related on a single particle level. In praxis, we cut the couplings to all other molecular orbitals and generate a transmission function containing the contribution of the HOMO only and then use $T_{HOMO}(E)=4V^{2}/((E-\epsilon_{HOMO})^{2}+4V^{2})$~\cite{troels}, where we obtain $T_{HOMO}(E)$ and $\epsilon_{HOMO}$ as direct results from NEGF-DFT and derive the transfer integral \textit{V} from a numerical fit. 

\end{subsubsection}

\begin{subsubsection}{Driving Force}

In principle the driving force $\Delta G^{0}$ in the electron transfer reaction we describe (where the respective Ru complex is neutral in the initial state and has a positive charge in the final state while the corresponding counter charge on the leads is assumed to be taken from the Fermi level of a metal surface with macroscopic dimensions) could be formulated by relating the ionisation potential (IP) of the complex as calculated from total energy differences of the charged and uncharged free molecule to the work function (WF) of the gold surfaces~\cite{nitzan6}. Such a definition, however, would neglect the effect of the adsorption of the molecule on the metal, i.e. Fermi level alignment and charge equilibrization~\cite{fermi1}-~\cite{fermi3}, since both the metallic WF and the molecular IP would be computed for the leads and molecule, separately. Therefore we use the HOMO's position relative to the Fermi level of the surface in the composite system~\cite{first} as a definition of $\Delta G^{0}$ instead, where the level alignment is accounted for correctly. As an additional benefit we can make a direct comparison between I/V curves for electron hopping calculated in this way and the transmission function for coherent tunneling, where both are derived on a single particle level and a gate potential can be applied in a rigid band approximation. 

\begin{table*}[tp]
 \begin{tabular}{|c|c|c|c|c|c|c|c|c|c|}
\hline
                &       G$_{coh}$       &    G$_{hop}$  & k$_{ox}$/k$_{red}$ &  V       & $\Delta G_0$ &        $\lambda_{in}$  &       $\lambda_{out}$ ($\lambda_{Born}/\lambda_{img}$)&       $\lambda_{tot}$   \\
\hline
 Ru1    &       2.0$\cdot$10$^{-5}$     &       1.2$\cdot$10$^{-25}$    &      3.1$\cdot$10$^{-12}$/2.4$\cdot$10$^{9}$&  1.35$\cdot$10$^{-3}$  & 1.250 &       0.177   &       0.421(0.661/-0.240)     &       0.597 \\
 Ru2    &       8.0$\cdot$10$^{-7}$     &       1.4$\cdot$10$^{-17}$    &      3.5$\cdot$10$^{-4}$/2.4$\cdot$10$^{8}$&  4.50$\cdot$10$^{-4}$   & 0.707 &       0.083   &       0.322(0.495/-0.172)     &       0.407 \\
 Ru3    &       1.6$\cdot$10$^{-8}$     &       5.4$\cdot$10$^{-16}$    &      1.3$\cdot$10$^{-2}$/6.5$\cdot$10$^{7}$&  8.03$\cdot$10$^{-5}$   & 0.576 &       0.059   &       0.315(0.446/-0.131)     &       0.374\\
\hline
\end{tabular}\protect\protect\caption{All quantities directly calculated from DFT for the three systems in Figure~\ref{fig.structures}, where G$_{coh}$ and G$_{hop}$ are given in units of G$_{0}$, and k$_{ox}$/k$_{red}$ in units of s$^{-1}$ while \textit{V}, $\Delta G^{0}$ and all contributions to $\lambda$ are presented in eV.}
\label{tab.data}
\end{table*}

\end{subsubsection}

\begin{subsubsection}{Reorganization Energy}

The total reorganization energy used in the equations~\ref{eqn.kox} and~\ref{eqn.kred} is defined as the sum of inner and outer contribution

\begin{equation}
\lambda_{tot}=\lambda_{in}+\lambda_{out}=\lambda_{in}+\lambda_{Born}+\lambda_{image},\label{lam_tot}
\end{equation}

where the latter can be further divided into a Born term accounting for the interaction of the charged molecule with the solvent and an image contribution, which describes the screening of the charge due to the vicinity of the metallic leads~\cite{ulstrup5}. The inner reorganization energy $\lambda_{in}$, i.e. the energy gain due to the relaxation of the nuclear positions of the molecule as a consequence of charging, can be calculated as the respective total energy difference of either the charged complex in its own equilibrium configuration minus a charged complex in the equilibrium configuration of the neutral molecule, or alternatively as total energy difference from two calculations where no charge is put on the two different equilibrium geometries. In praxis, we take the average of these two possibilities.

For $\lambda_{Born}$, we employ a solvent continuum model as already suggested by Marcus~\cite{marcus1}-~\cite{marcus3}, who used the Born approximation for calculating the solvation energy of spherical ions~\cite{born}. For the Ru complexes in our article, we need to extend this to the generalized Born approximation (GBA)~\cite{still}

\begin{equation}
\lambda_{Born}=(\frac{1}{\epsilon_{\infty}}-\frac{1}{\epsilon_{s}})\sum_{i,j}^{N}\frac{\Delta q_{i}\cdot\Delta q_{j}}{f_{GB}},\label{lam_born}
\end{equation}

where $\epsilon_{\infty}$ and $\epsilon_{s}$, are the optical and static permittivity of the solvent, respectively, while $\Delta q_{i,j}$ are the partial charge differences between the neutral and the oxidized state of the free molecule in vacuum, which were calculated as Mulliken charges from DFT and where the van der Waals radii entering $f_{GB}$ according to Ref.~\cite{still} have been taken from Ref.~\cite{jorgensen}. 

\end{subsubsection}

\begin{subsubsection}{Screening by the Leads}

Within an image charge model, the contribution of the screening of the charge on the molecule between two planar metal surfaces to the reorganization energy can be described by an infinite sum of Coulomb interactions between the partial charges on the molecule and their infinite number of image charges in the electrodes~\cite{ulstrup5,kristian1,screening}

\begin{multline}
\lambda_{image}=-\frac{1}{2}(\frac{1}{\epsilon_{\infty}}-\frac{1}{\epsilon_{s}})\sum_{i,j}^{N}\Delta q_{i}\cdot\Delta q_{j}\cdot\\
\sum_{n=1}^{\infty}\Biggl[\frac{1}{\sqrt{(z_{i}+z_{j}-2nL)^{2}+R_{ij}^{2}}}-\frac{2}{\sqrt{(z_{i}-z_{j}+2nL)^{2}+R_{ij}^{2}}}+\\
\frac{1}{\sqrt{(z_{i}+z_{j}+2(n+1)L)^{2}+R_{ij}^{2}}}\Biggr],
\label{lam_img}
\end{multline}

where $R_{ij}^{2}=(x_{i}-x_{j})^{2}+(y_{i}-y_{j})^{2}$ and x$_{i,j}$, y$_{i,j}$, z$_{i,j}$ are the positions of the atoms of the molecule, with the z-direction being the transport direction. 

\end{subsubsection}
\end{subsection}
\end{section}

\begin{section}{Results and Discussion}

\begin{subsection}{Direct Comparison of the Conductance from Coherent Tunneling and Electron Hopping}

In Table~\ref{tab.data} we show all results we derived for the structures in Figure~\ref{fig.structures} directly from DFT calculations, i.e. the conductances G$_{coh}$ and G$_{hop}$ for coherent tunneling and electron hopping for zero bias and zero gate voltage, respectively, as well as the values for k$_{ox}$/k$_{red}$, \textit{V}, $\Delta G^{0}$ and $\lambda$ which define G$_{hop}$ through Equations~\ref{eqn.kox}-~\ref{eqn.g_hop}.  While G$_{coh}$ decays exponentially with the length of the molecule as expected, G$_{hop}$ shows an increase with molecular length at least for the three molecules under investigation. The transfer integral \textit{V} decreases as the amplitude of the HOMO at the anchor group diminishes with a rise in molecular size. The reorganization energy also decreases with the size of the molecule because both the relaxations of internal degrees of nuclear freedom and the polarization of the solvent become energetically easier for a larger molecular volume. Finally, also the driving force $\Delta G^{0}$ decreases with the molecular length, because the HOMO-LUMO gap becomes smaller with the length of a semiconducting wire and therefore the HOMO moves closer to E$_{F}$.

\begin{figure}
\includegraphics[width=1.0\linewidth,angle=0]{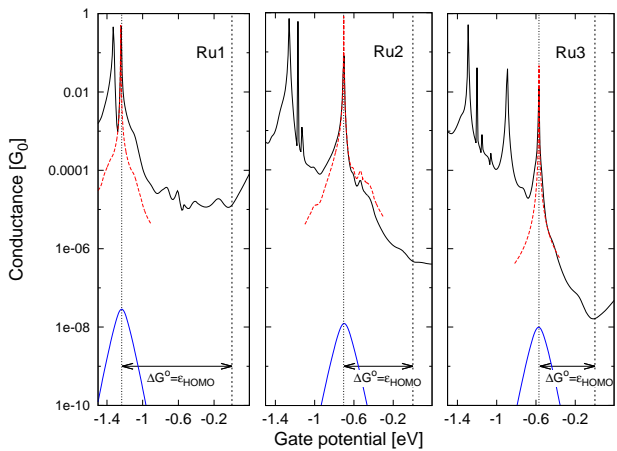}
\caption[cap.Pt6]{\label{fig.transmission}The transmission functions T(E) for coherent tunneling for the three structures in Figure~\ref{fig.structures} (black lines) are directly compared with the voltage dependent behaviour of G$_{hop}$ (blue lines) in dependence on an electrochemical gate, where the respective voltage is simply added to $\Delta G^{0}$ in Equations~\ref{eqn.kox} and~\ref{eqn.kred} as an overpotential. The dotted red line shows T(E) for electrons mediated only by the HOMO as described in detail in the main text.}
\end{figure}

\begin{figure}
\includegraphics[width=1.0\linewidth,angle=0]{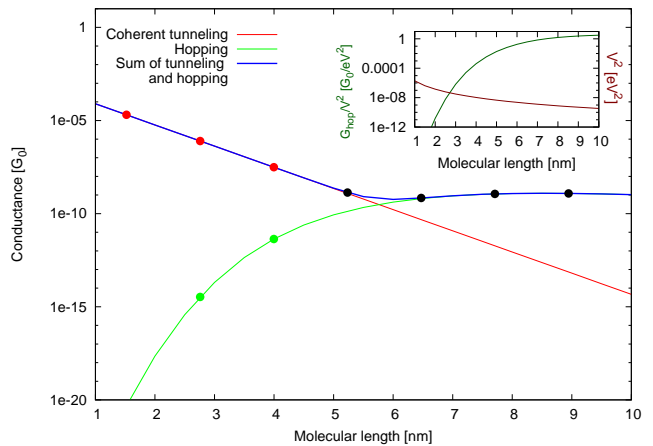}
\caption[cap.Pt6]{\label{fig.extrapol}Extrapolation of the molecular length dependence of G$_{coh}$ (red line) with an exponential fit, and the length dependence of \textit{V}, $\lambda$ and $\Delta G^{0}$, where the first two quantities were fitted with power laws and the third one with an exponential function, for the definition of G$_{hop}$ (green line). The blue line represents the sum G$_{coh}$+G$_{hop}$, which corresponds to the conductance accessible to experiments for this type of molecular wire. The colored dots on each line refer to the values directly calculated from DFT for Ru1, Ru2 and Ru3, respectively, while at all higher lengths the extrapolations have been taken and marked with black dots. The inset shows the respective length dependent evolutions of the contributions to G$_{hop}$ coming from \textit{V$^{2}$} (red line) and the remaining factor (green line), which is exclusively defined by $\Delta G^{0}$ and $\lambda$ and is obtained by dividing G$_{hop}$ through \textit{V$^{2}$}.}
\end{figure}

The transfer rates k$_{ox}$ and k$_{red}$ in Table~\ref{tab.data} are completely defined by the three parameters \textit{V}, $\lambda$ and $\Delta G^{0}$ through Eqn. \ref{eqn.kox} and \ref{eqn.kred}, and in their dependence on the gate voltage behave like error functions where k$_{red}$ increases when k$_{ox}$ decreases with a crossing point at $\Delta G^{0}$. Therefore a product of the transfer rates as in Eq. \ref{eqn.g_hop} results in a peak around $\Delta G^{0}$ since one of the two factors is always minimal at larger energetic distances from $\Delta G^{0}$. A reduction of $\Delta G^{0}$ with the length of the molecule leads to a shift of this peak towards the Fermi Level resulting in an increase of G$_{hop}$ at zero gate voltage. An increase of the reorganization energy $\lambda$ on the other side results in a widening and lowering of this peak because it leads to a shift of k$_{ox}$ and k$_{red}$ in opposite directions in their dependence on the gate voltage.

In Figure~\ref{fig.transmission} we directly compare the transmission function T(E) for coherent tunneling as obtained from NEGF-DFT with the hopping conductance as a function of a gate voltage or overpotential. While for the former it is assumed that V$_{gate}$ is equal to the kinetic energy of incoming electrons E, for the latter, a zero gate voltage means that $\Delta G^{0}$=-$\epsilon_{HOMO}$ as obtained by a subdiagonalization procedure from the transport Hamiltonian of the composite junction, where for a finite voltage the applied potential is just added to $\Delta G^{0}$ in Equations~\ref{eqn.kox} and~\ref{eqn.kred} as a scalar. Both assumptions are just implementations of a rigid band approximation within a single particle picture. While for T(E) the transmission peak corresponding to the HOMO (red dotted line) and slightly offset due to hybridization effects is moving ever closer to E$_{F}$ with an increasing length of the molecule, it also becomes narrower since \textit{V} is decreasing at the same time where the accumulated effect is the exponential decay of G$_{coh}$. The blue peak illustrating the gate voltage dependence of G$_{hop}$ is also moving towards the Fermi level with an increase in molecular size where its maximum is always at the energetic position of $\epsilon_{HOMO}$=$\Delta G^{0}$, representing zero overpotential. Its width and height on the other hand are defined by $\lambda$ and \textit{V}, and a continuous rise in the zero bias and zero gate G$_{hop}$ is found when moving from Ru1 to Ru3. 

\end{subsection}

\begin{subsection}{Dependence of the Conductance on Molecular Length and Crossover Point between the two Transport Regimes}

In Table~\ref{tab.data} it can also be seen, that although the values of G$_{coh}$ and G$_{hop}$ approach each other when going from Ru1 to Ru3, no crossover point between the two regimes can be reached within the scope of these three molecules. Since the junction with the Ru3-complex in Figure~\ref{fig.structures} defines about the limit of what can still be calculated with DFT in terms of the computational costs, we used exponential fits for G$_{coh}$ and $\Delta G^{0}$ as well as power law fits for \textit{V} and $\lambda$, for an extrapolation of G$_{coh}$, G$_{hop}$ and G$_{coh}$+G$_{hop}$, where the results for the wire length ranging up to 10 nm are shown in Figure~\ref{fig.extrapol}. 

The exponential decay for G$_{coh}$ is a well-known property of coherent tunneling where we evaluated a decay constant $\beta$ of 2.7 nm$^{-1}$ which matches well with what is usually found in experiments for conjugated molecular wires~\cite{wandlowski1}. For the length dependence of G$_{hop}$ we find an increase up to 6 Ru centers but the roughly linear decay expected for Ohmic transport sets in for molecular wires longer than that. 

\begin{figure*}
\includegraphics[width=1.0\linewidth,angle=0]{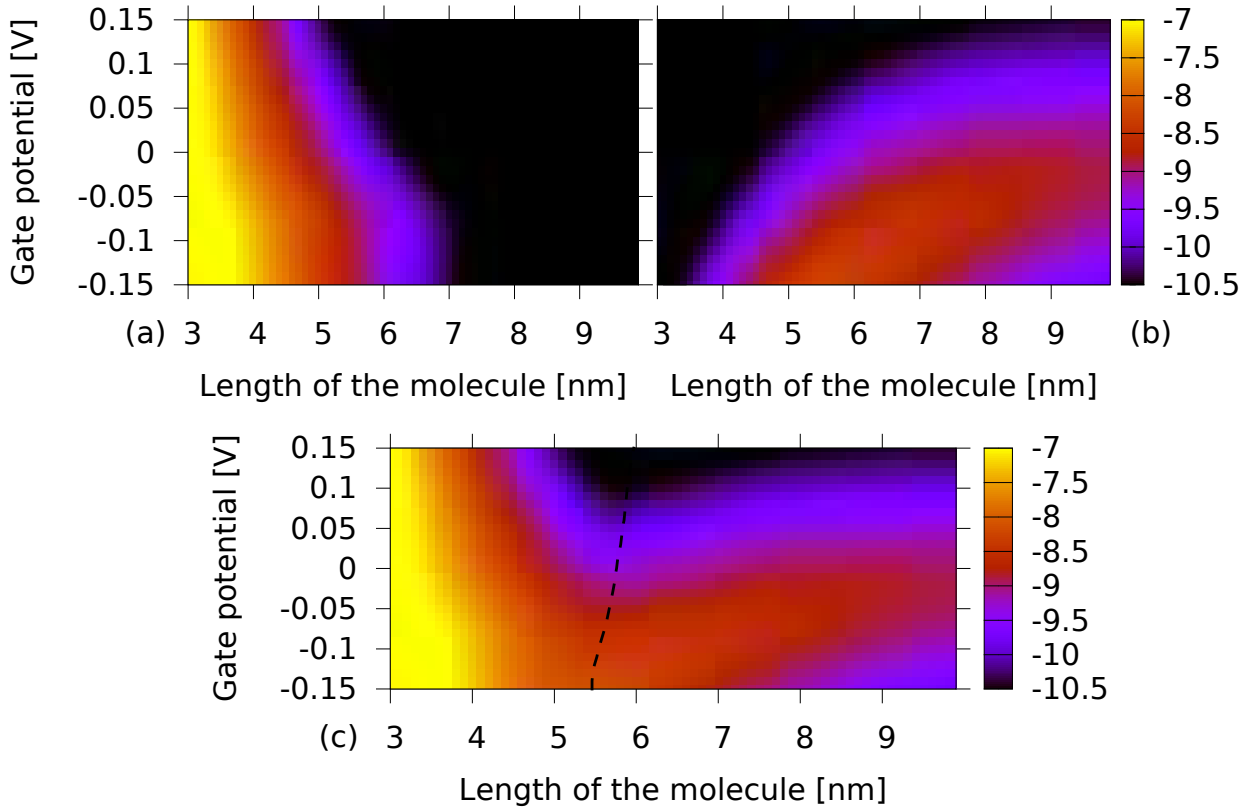}
\protect\protect\caption[cap.Pt6]{\label{fig.gate}Color contours for the two-dimensional dependence of a) G$_{coh}$, b) G$_{hop}$ and c) G$_{coh}$+G$_{hop}$ on the molecular length and on an electrochemical gate potential, where the color code is given on the side of the panels with the numbers referring to x in 10$^{x}$. In panel c) a black dashed line is drawn where G$_{coh}$=G$_{hop}$.}
\end{figure*}

The particular functions we used for the fitting of V, $\lambda$ and $\Delta$G$^0$ we chose because they gave the best representation of our data but their overall behaviour has an intuitive physical explanation. At very small molecular lengths (the first two points in Fig.~\ref{fig.extrapol}) both $\Delta$G$^0$ and $\lambda$ are rather large and since they enter the exponents in Eq.~\ref{eqn.kox} and ~\ref{eqn.kred} with a negative sign, this results in small values of k$_{ox}$, k$_{red}$ and G$_{hop}$. The larger the molecular length the smaller $\lambda$ and $\Delta$G$^0$ become and therefore the larger G$_{hop}$. The reason for the drastic increase in G$_{hop}$ and its subsequent stabilization is that we define $\Delta$G$^0$ as the distance of the HOMO to the Fermi level which becomes smaller with the molecular length (as can be seen from the peak shift in Fig.~\ref{fig.transmission}) and then converges to zero asymptotically. The behaviour of $\lambda$ points in the same direction where the screening by the solvent decreases if a charge of one electron is spread out over larger molecular volumes and this effect also has an asymptotic limit. At some point the continuous decrease of V$^2$ in Eqns. ~\ref{eqn.kox} and ~\ref{eqn.kred} determines the further length dependence of G$_{hop}$.

In the inset of Figure~\ref{fig.extrapol} we show the two factors separately, which determine G$_{hop}$ in Equations~\ref{eqn.kox} and~\ref{eqn.kred} where \textit{V$^{2}$} decays with 1/length (red line) and the exponential containing $\Delta G^{0}$ and $\lambda$ (green line) rises strongly for small lengths but then approaches 1 at 8-10 nm. 

From Figure~\ref{fig.extrapol} we can identify the crossover or transition point from coherent transport to electron hopping where G$_{coh}$=G$_{hop}$ and find it at a molecular length of 5.76 nm. This is in the same range as the 5.6-6.8 nm for polythiophenes~\cite{yamada} and 5.2-7.3 nm for oligofluorene based molecular wires~\cite{hines} found in recent experiments but somewhat higher than the $\sim$3 nm for pyridil-terminated OAEs~\cite{wandlowski1}, the $\sim$2.75 nm for a series of amine-terminated OPEs~\cite{wang1} and the $\sim$4 nm for thiol-anchored oligophenyleneimines~\cite{frisbie1} which were also recently measured. 

\end{subsection}

\begin{subsection}{Application of an Electrochemical Gate Potential}

Finally, we discuss the dependence of G$_{coh}$, G$_{hop}$ and G$_{coh}$+G$_{hop}$ on an electrochemically applied gate potential as depicted in Figure~\ref{fig.gate}a, b and c, respectively, where we chose a relatively small range of potentials because we use a rigid band approximation for both transport regimes which assumes that the electronic structure is undisturbed by the applied voltage. In Ref.~\cite{wandlowski2} it was found experimentally that increasing the gate voltage in one direction leads to a rapid increase in the conducance while a voltage with the opposite sign had no effect. This was interpreted as a reduction of the complex with a negative potential but in terms of our Figure~\ref{fig.transmission} it can also be seen as a climbing up of the peak related to the HOMO in both transport regimes. As pointed out earlier there is a marked difference between the two regimes, where the peak is rather narrow for coherent transport and the range of voltages we chose is not sufficient for generating any clear trends in Figure~\ref{fig.gate}a but a negative voltage distincly increases G$_{hop}$ in Figure~\ref{fig.gate}b, which reflects the broader peak for hopping found in Figure~\ref{fig.transmission}. In Figure~\ref{fig.gate}c we also drew a black line where G$_{coh}$=G$_{hop}$ and find that the transition point between coherent tunneling and hopping is moving to smaller wire lengths for negative potentials in this two-dimensional picture. This finding provides additional means for experiments to shift the length range of Ohmic behaviour towards smaller molecules. In many cases this could make the transition point accessible for experimental studies when longer wires are hard to synthesize or difficult to handle in measurements. 

\end{subsection}
\end{section}

\begin{section}{Summary}\label{sec:summary}

We performed DFT calculations for the conductance of a series of single molecule junctions with redox-active molecular wires containing one, two and three Ru atoms, where we treated both coherent tunneling and electron hopping within the same single particle framework, which allowed for a direct quantitative comparison of the two electron transport regimes. An extrapolation of our ab initio results made it possible to identify a molecular length for a transition point between them at 5.76 nm, which is rather close to the values reported from measurements on similar wires. We also investigated the dependence of this transition length on an electrochemically applied gate voltage, where we found that it can be shifted by an external potential which provides experimentalists with another tool to study the crossover between transport regimes. This is also of technological relevance because only hopping has the Ohmic length dependence required for wire applications and according to our finding a gate voltage can move its onset towards shorter wires.

\end{section}


\begin{acknowledgments}

G.K. and R.S. are currently supported by the Austrian Science Fund FWF, project Nr. P22548. We are deeply indebted to the Vienna Scientific Cluster VSC, on whose computing facilities all calculations presented in this article have been performed (project Nr. 70174) and where we were provided with extensive installation and mathematical library support by Markus St\"{o}hr and Jan Zabloudil in particular. We gratefully acknowledge helpful discussions with Tim Albrecht, Michael Inkpen and Thomas Wandlowski.
\end{acknowledgments}


\bibliographystyle{apsrev}

\expandafter\ifx\csname natexlab\endcsname\relax\global\long\def\natexlab#1{#1}
 \fi \expandafter\ifx\csname bibnamefont\endcsname\relax \global\long\def\bibnamefont#1{#1}
 \fi \expandafter\ifx\csname bibfnamefont\endcsname\relax \global\long\def\bibfnamefont#1{#1}
 \fi \expandafter\ifx\csname citenamefont\endcsname\relax \global\long\def\citenamefont#1{#1}
 \fi \expandafter\ifx\csname url\endcsname\relax \global\long\def\url#1{\texttt{#1}}
 \fi \expandafter\ifx\csname urlprefix\endcsname\relax\global\long\def\urlprefix{URL }
 \fi \providecommand{\bibinfo}[2]{#2} \providecommand{\eprint}[2][]{\url{#2}} 

\end{document}